\documentclass[a4paper,12pt,fleqn]{article}
\newcommand{\be}{\begin{eqnarray}&&}
\usepackage{graphicx}
\usepackage{appendix}
\usepackage{wrapfig}
\usepackage{amsmath}
\usepackage{amsfonts}
\usepackage{appendix}
\usepackage[left=3cm,top=3cm,bottom=3cm,right=3cm]{geometry}

\begin{document}
\title{Role of $\phi$ decays for $K^-$ yields\\
in relativistic heavy-ion collisions}
\author{H.~Schade$^{1,2}$, Gy.~Wolf$^{3}$, B.~K\"ampfer$^{1,2}$}
\maketitle
\begin{center}
 \textit{$^{1}$Forschungszentrum Dresden-Rossendorf, PF 510119, 01314 Dresden, Germany\\
$^{2}$TU Dresden, Institut f\"ur Theoretische Physik, 01062 Dresden, Germany\\
$^{3}$KFKI RMKI, H-1525 Budapest, POB 49, Hungary}
\end{center}

\begin{abstract}

The production of strange mesons in collisions of Ar+KCl at a kinetic beam energy of 1.756 AGeV is studied within a transport model of Boltzmann-\"Uhling-Uhlenbeck (BUU) type. In particular,  $\phi, K^+$ and $K^-$ yields and spectra are compared to the data mesured recently by the HADES collaboration and the $\phi$ yield measured previously by the FOPI collaboration. Our results are in agreement with these data thus presenting an interpretation of the subleading role of $\phi$ decays into $K^-$'s and confirming the importance of the strangeness-exchange channels for $K^-$ production. 

\end{abstract}

\section {Introduction}

Strange hadrons produced in heavy-ion collisions at beam energies in the 1-2 AGeV range have been considered as useful probes of the reaction dynamics, the equation of state as well as medium modifications \cite{Hartnack:2007wu, Hartnack:2005tr, Hartnack:2001zs, Hartnack:2001zs_err, Hartnack:2003dt}. The distinctly different behavior of $K^+$ and $K^-$ mesons can be attributed, to some extent, to repulsive and attractive potentials by the surrounding nuclear matter. These potentials may be translated into effective mass shifts, often advocated values are +20 MeV for $K^+$ and -75 MeV for $K^-$ at nuclear saturation density, respectively \cite{Hartnack:2001zs, Hartnack:2001zs_err, Kolomeitsev:1995xz, Waas:1996fy, Tsushima:1997df, SchaffnerBielich:1999cp, Tolos:2002ud, Tolos:2000fj, Brown:2000vt, Brown:2001wq, Li:1996fw, Li:1994vd, Li:1994cu, Li:1997zb, Ramos:1999ku, Bratkovskaya:1997dh, Cassing:2003nz}. Proton-induced $K^\pm$ production at nuclei supports experimentally this picture \cite{Scheinast:2005xs}, while detailed coupled channel calculations \cite{Lutz:1997wt} point to more involved modifications of $K^\pm$ spectral functions, in particular for $K^-$. At the mentioned beam energies, the production of $K^-$ is below the free $NN$ threshold. In so far, $K^-$ yields are expected to depend sensibly on the details of the network of hadron collisions, the $K^-N$ potential and the mean field in the course of heavy-ion collisions.\\
The strangeness transfer reactions $\pi Y \rightarrow NK^-$ $(Y = \Lambda, \Sigma)$ and $BY \rightarrow NNK^-$ $(B = N, \Delta)$ has been identified in transport models as dominating processes yielding most of the finally observed $K^-$ \cite{Hartnack:2007wu, Hartnack:2005tr, Hartnack:2003dt}. While a fairly large set of binary hadronic reactions is included in many transport simulation codes \cite{Kolomeitsev:2004np, Fuchs:2005zg, Hartnack:2005eh, Cassing:1999es}, the role of the decay $\phi \rightarrow K^+K^-$ (branching ratio 49.2 \%) for creating $K^-$ has not yet explored exhaustively. In fact, the experimental information on $\phi$ was scarce until recently: The early data samples from FOPI \cite{Mangiarotti:2003es, Kotte_Hirschegg} allowed the identification of $\phi$'s via the invariant mass distribution of $K^+ - K^-$ pairs in restricted phase-space regions for the reactions Ni (1.93 AGeV) + Ni and Ru (1.69 AGeV) + Ru. The extrapolation to the total yield is hampered by uncertainties since spectra could not be deduced due to the restricted statistics.\\
The situation now changed as the HADES collaboration published for the reaction Ar (1.756 AGeV) + KCl a multiplicity ratio $\phi/K^- = 0.37 \pm 0.13$, and a transverse momentum spectrum of $\phi$ is available for the rapidity bin $y_{lab} = 0.2-0.6$ \cite{Agakishiev:2009ar}. Equipped with these new data one can now try to answer the question \cite{Kampfer:2001mc} whether the $\phi$ decay channel competes with the strangeness transfer reactions as source of $K^-$. In doing so, we employ a transport code of BUU type \cite{Kolomeitsev:2004np}. This code has been used to analyze the previous FOPI data \cite{Mangiarotti:2003es, Kotte_Hirschegg} in \cite{Barz:2001am} and to predict $\phi$ yields in proton-nucleus reactions \cite{Barz:2003wz}. \\
While $\phi$ mesons have net strangeness zero, they have a dominating $s\bar{s}$ component. This qualifies them as some type of multi-strange probes of compressed nuclear matter. Clearly, they differ from real multi-strange hadrons such as $\Xi$. Given the fact that a first $\Xi^-$ $(dss)$ signal has been detected in subthreshold heavy-ion collisions \cite{Agakishiev:2009rr} with a yield above predictions of the otherwise so successful thermo-statistical models \cite{Andronic:2003zv, Andronic:2005yp}, it may be interesting to study the $s\bar{s}$ channel in this context.\\
In elementary hadron collisions, in particular proton-proton collisions, the analysis of $\phi$ production (see for instance \cite{Titov:2000bn}) is related to the OZI rule and the nucleon-$\phi$ coupling as well as the issue of nucleon resonances strongly coupling to $\phi$. Even bound states of the $\phi$-$N$ system are under consideration \cite{Liska:2007de}. Valuable experimental information on $\phi$ production near the threshold has been obtained in recent measurements by COSY-ANKE \cite{Hartmann:2006zc} including the quasi-free $pn$ reaction \cite{Maeda:2006wv}. For further investigations of these issues we refer the interested reader to \cite{Kaptari:2004sd, Kaptari:2008nb, Sibirtsev:2005zc}. Photon-induced $\phi$ production at nuclei has been studied experimentally by CLAS \cite{Djalali:2008zza} with the result of finding a sizeable broadening \cite{Djalali:2008zza}. Proton-induced $\phi$ production has been considered theoretically with predictions of a non-trivial atomic mass number dependence \cite{Paryev:2008ck, Barz:2003dd, Sibirtsev:2006yk}.\\
QCD sum rule studies \cite{Kuwabara:1995ms, Hatsuda:1992bv, Hatsuda:1991ez, Zschocke:2002mn, Kampfer:2002pj} predict the possibility of in-medium modifications of $\phi$ mesons which may be condensed in an effective mass. (Interestingly, the strength of the in-medium modification depends on the hidden strangeness content of the nucleon.) We are going to include such a modification via a weakly attractive potential, parametrized similarly to anti-kaons.\\
The various hadronic models \cite{Oset:2000eg, Cabrera:2002hc, Klingl:1997tm, Asakawa:1994tp, Song:1996gw, AlvarezRuso:2002ib, Blaizot:1991af, Bi:1990pp, Holt:2004tp} address further the in-medium modifications of $\phi$ mesons, where \cite{Bi:1990pp, Holt:2004tp} focus especially on the in-medium width. Sizeable broadening effects are found in \cite{Vujanovic:2009wr} from the foreward scattering amplitude which result in a tiny shift of the spectral $\phi$ peak. Experimentally, the KEK-PS-E325 collaboration announced a first indication of in-medium modifications of the $\phi$ meson in \cite{Muto:2005za, Muto:2006eg} in already mentioned proton-induced reactions at nuclei. \\
In ultrarelativistic heavy-ion collisions, the $\phi$ meson is expected to decouple earlier from the bulk of excited strongly interacting matter \cite{Abelev:2008zk} due to its small interaction cross section and thus may allow for a differential diagnostics of the freeze-out \cite{Hirano:2007ei, Santini:2006cm}.\\
As the $\phi$ production is suppressed in elementary hadronic reactions because of the OZI rule \cite{OZI}, the so-called catalytic reactions $\pi Y \rightarrow \phi Y$ and $\bar{K}N \rightarrow \phi Y$ have been proposed in \cite{Kolomeitsev:2009yn} as potentially strong sources in relativistic heavy-ion collisions. We include a fairly large set of hadronic reactions where a $\phi$ meson is involved. This will set some bound on the importance of these poorely known catalytic channels. \\
Our paper is organized as follows. In Section 2, we compare $K^+$, $K^-$ and $\phi$ spectra calculated within our transport model approach with the data \cite{Agakishiev:2009ar}. The detailed analysis of $\phi, K^+$ and $K^-$ channels is performed in Section 3. We address thereby the time evolution, freeze-out conditions, centrality dependence, the role of individual channels, effects of the equation of state, in-medium masses, and compare with previous FOPI data. Our summary can be found in Section 4. The utilized cross sections and further employed parameters of the transport code are described in the Appendix.

\section {$K^\pm$ and $\phi$ spectra}

With the settings of the transport code listed in the Appendix we obtain the transverse momentum spectra and rapidity distributions of $K^\pm$ and $\phi$ for the collision of Ar (1.756 AGeV) + KCl exhibited in Figs. 1 - 6.\\
\begin{figure}
\begin{center}
\centering \includegraphics[width=8cm]{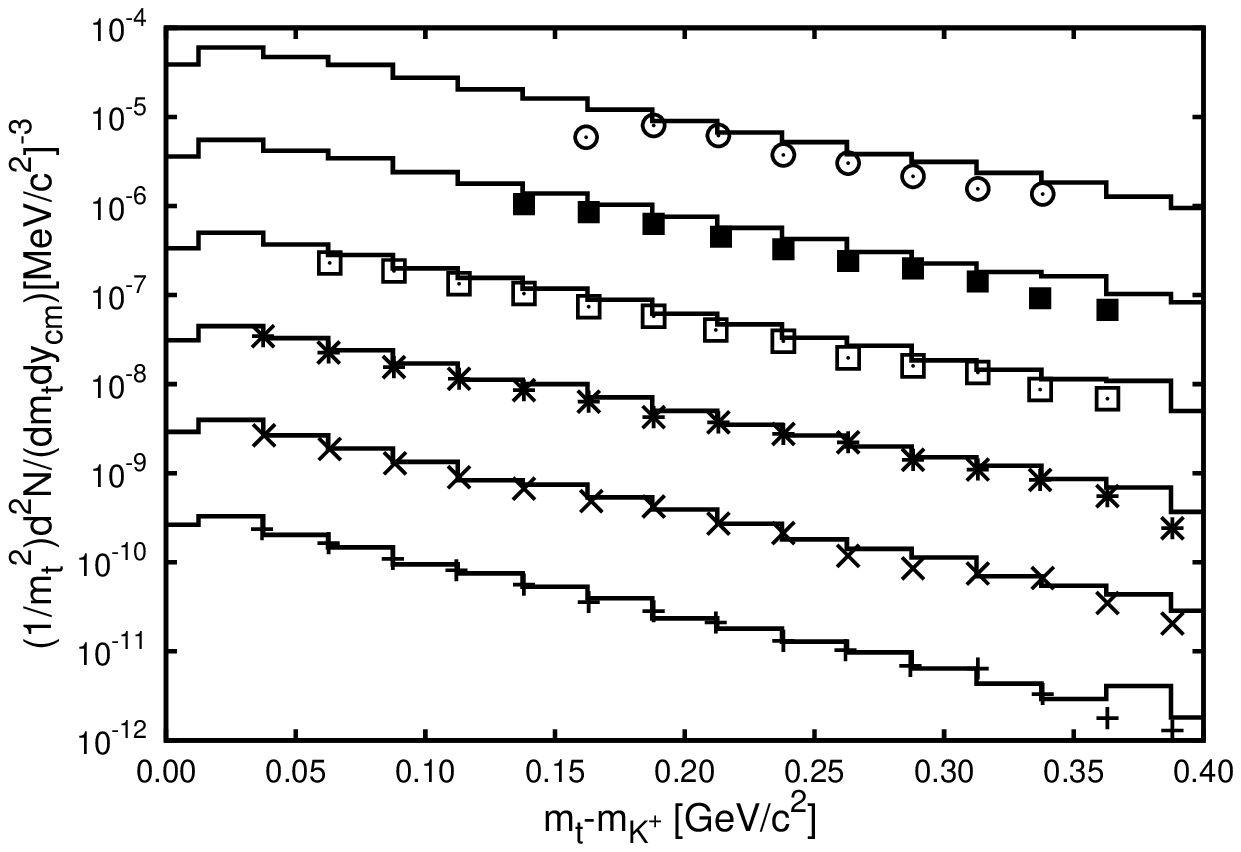}
\caption{Transverse-mass spectra of $K^+$ mesons for six different rapidity bins (from bottom to top: $0.1 < y_{lab} < 0.2$ to $0.6 < y_{lab} < 0.7$ with scaling factors from $10^0$ to $10^5$). The solid histograms are for an impact parameter of $b = 3.9$ fm. Data (symbols) are from \cite{Agakishiev:2009ar}.} 
\label{fig:mtkplus}
\vspace*{0.5cm}
\centering \includegraphics[width=8cm]{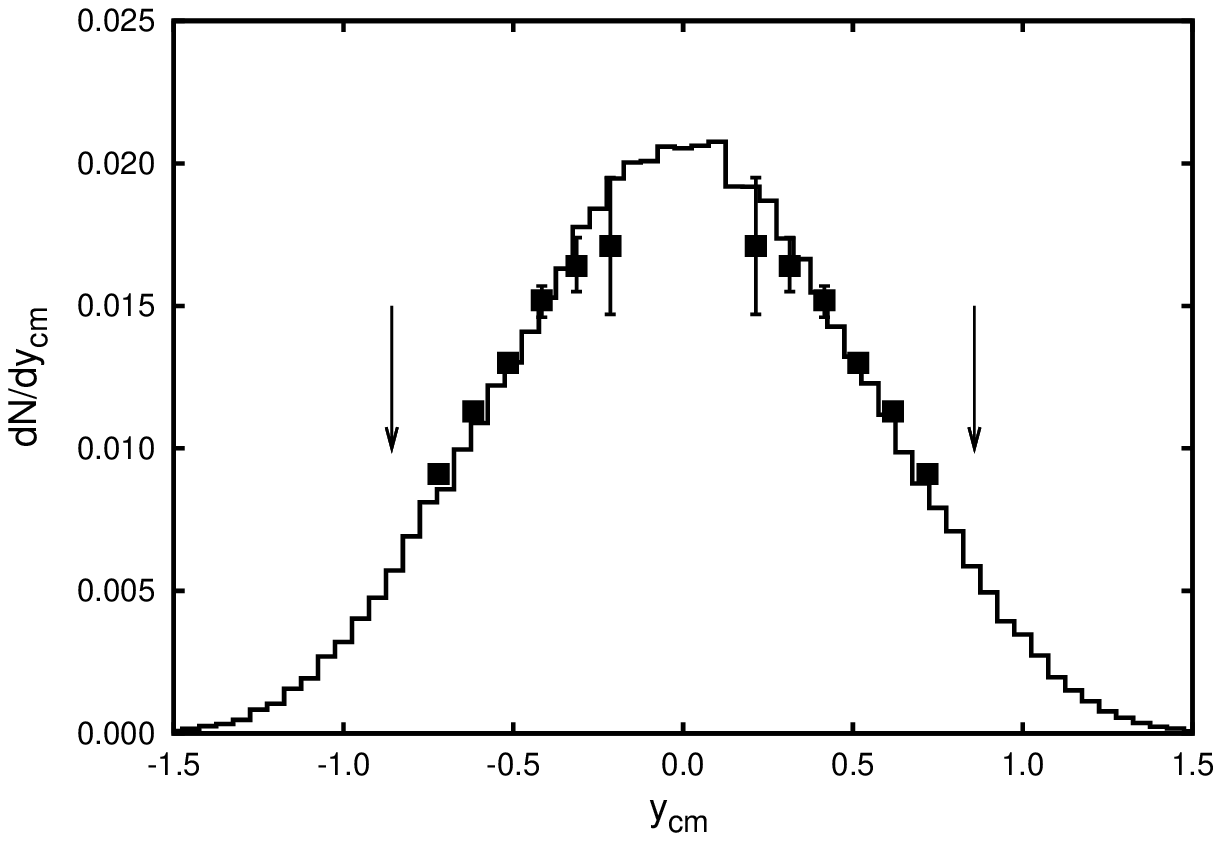}
\caption{Rapidity distribution of $K^+$ mesons in the c.m.~frame. The solid histogram represents a simulation with impact parameter $b = 3.9$ fm. Arrows indicate the target- and projectile rapidities of $\pm$ 0.858. Data are from \cite{Agakishiev:2009ar}.} 
\label{fig:rapkplus}
\end{center}
\end{figure}
Figure 1 exhibits the transverse mass ($m_t = \sqrt{m_K^2 + {\vec p}_t^{\;2}}$, where $p_t$ is the transverse momentum) spectra for various rapidity bins for $K^+$ mesons. One recognizes a fairly good agreement with the data \cite{Agakishiev:2009ar} for an impact parameter of $b = 3.9$~fm (solid histogram) with a tendency to overestimate the data at midrapidity. Nevertheless, this impact parameter describes best the rapidity distribution (see next paragraph). The $m_t$ or $p_t$ data do not allow for a tight constrain of the $K^+$ potential.\\
\begin{figure}
\begin{center}
\centering \includegraphics[width=8cm]{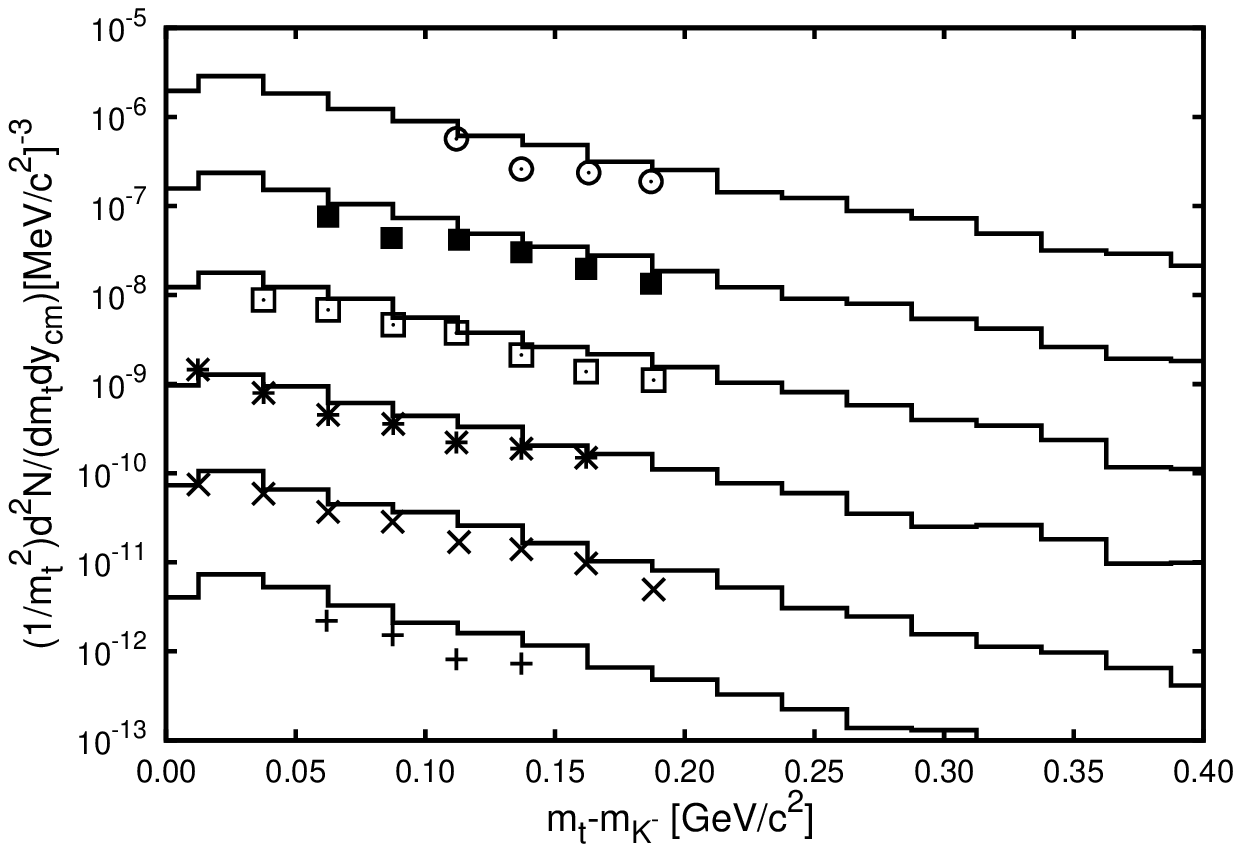}
\caption{The same as in Fig.~\ref{fig:mtkplus} but for $K^-$ mesons.}
\label{fig:mtkminus}
\vspace*{0.5cm}
\centering \includegraphics[width=8cm]{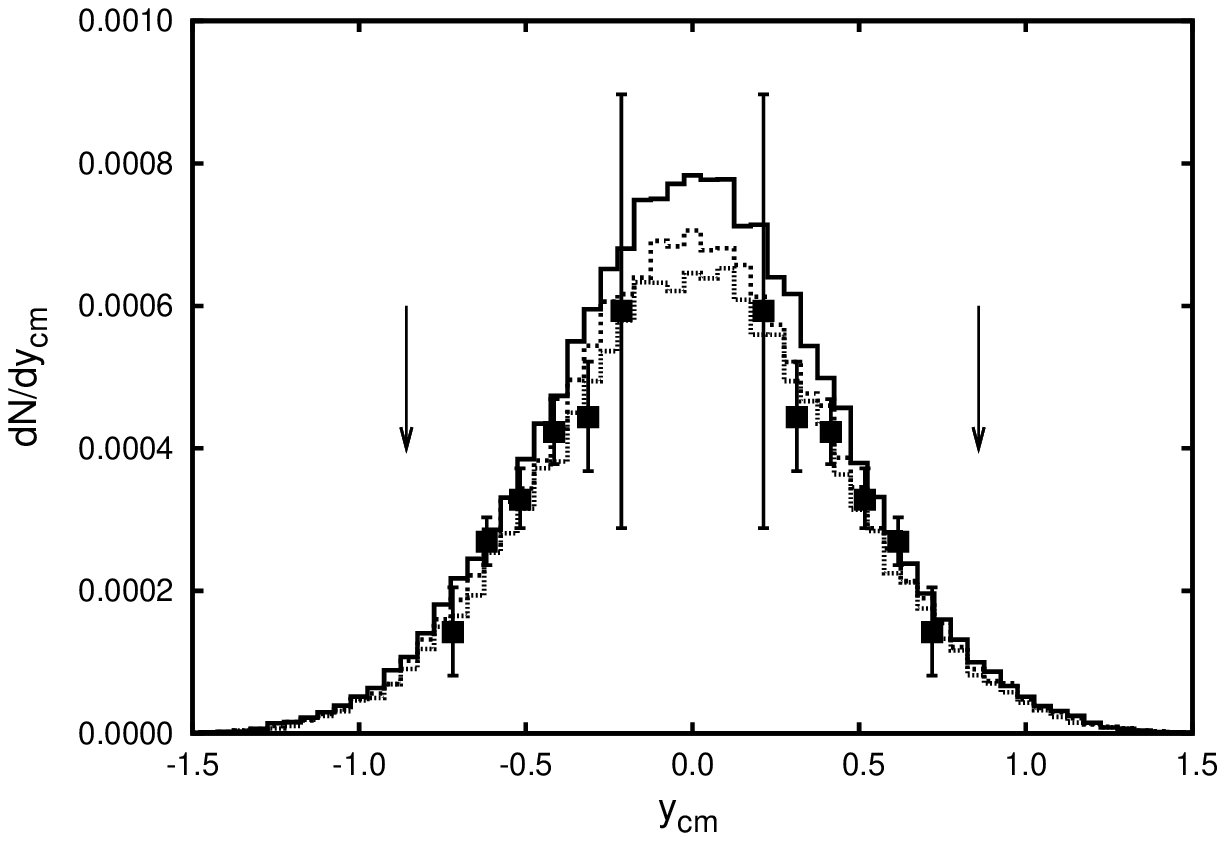}
\caption{The same as in Fig.~\ref{fig:rapkplus} but for $K^-$ mesons. Exhibited are simulations with nuclear incompressibility parameter $\kappa$ = 215~MeV (default), 290~MeV, and 380 MeV (solid, dashed and dotted histograms from top to bottom) with the resulting yields $7.8 \times 10^{-4}$, $7.4 \times 10^{-4}$, and $6.9 \times 10^{-4}$, respectively.}
\label{fig:rapkminus}
\end{center}
\end{figure}
Figure 2 compares the rapidity distribution with the data \cite{Agakishiev:2009ar}. The given linear scale is sensitive to small differences which are not visuable in the transverse mass spectra due to the used logarithmical scale. For instance, if we mimic the level-1 (LVL1) trigger, employed in the experiment \cite{Agakishiev:2009ar}, with an impact parameter distribution according to \cite{Krizek:2008} we find a 20 \% overestimate of the data. The shape of the rapidity distribution is the same for an impact parameter of $b = 3.6$ fm which is the mean value of the impact parameter distribution in \cite{Krizek:2008}. This fact confirms the assumption of an approximate impact parameter independence of the various particle production channels in the given comparatively small collision system (see section 3.3 for further details). An 8 \% increase of the impact parameter to 3.9 fm results in an optimum description of the data, see Fig.~\ref{fig:rapkplus}. The $K^+$ multiplicity deduced from our calculation exhibited in Fig.~\ref{fig:rapkplus} is $2.7 \times 10^{-2}$ to be compared with $(2.8 \pm 0.4) \times 10^{-2}$ reported in \cite{Agakishiev:2009ar}. We keep this impact parameter for the subsequent calculations, except in subsection 3.3 where the centrality dependence will be studied.\\
\begin{figure}
\begin{center}
\centering \includegraphics[width=8cm]{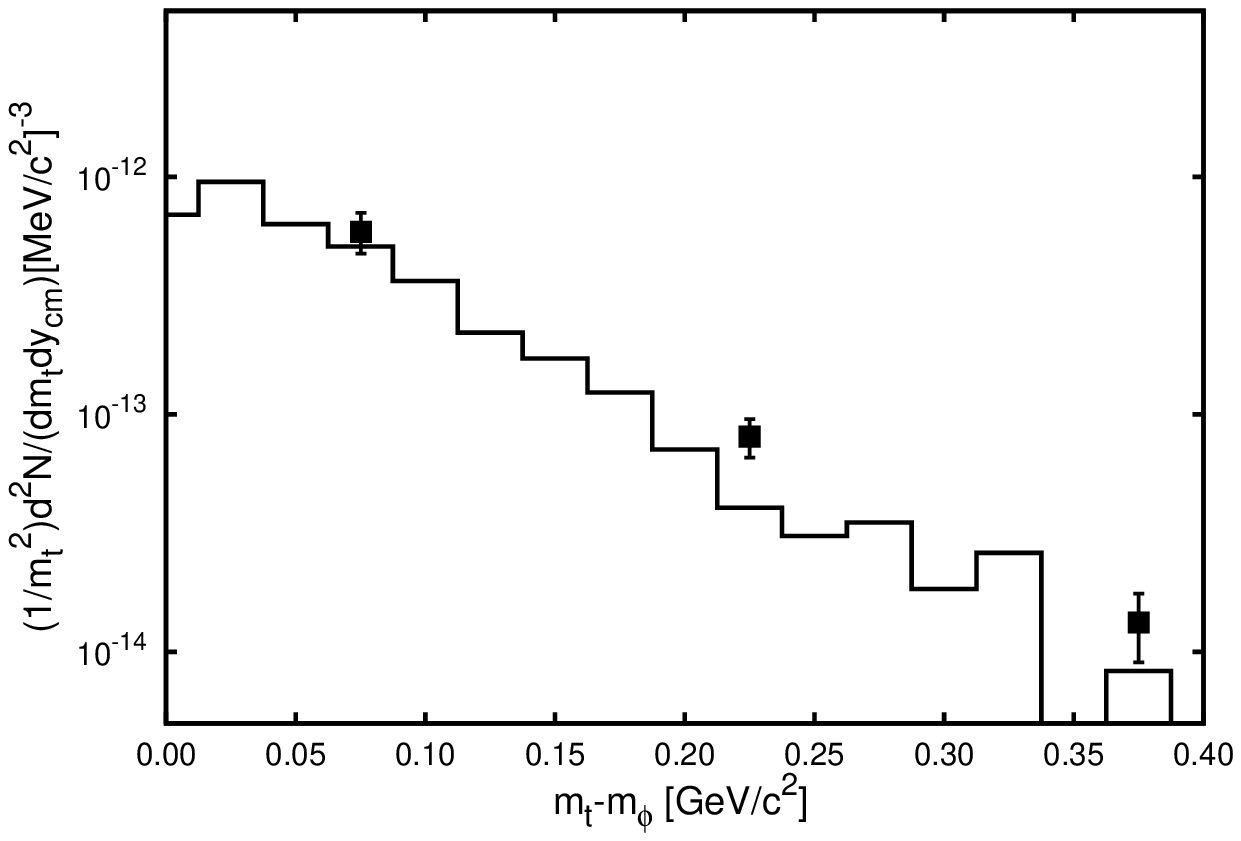}
\caption{The same as in Fig.~\ref{fig:mtkplus} but for $\phi$ mesons in the rapidity bin $0.2 < y_{lab} < 0.6$. Data are from \cite{Agakishiev:2009ar}.} 
\label{fig:mtphi}
\vspace*{0.5cm}
\centering \includegraphics[width=8cm]{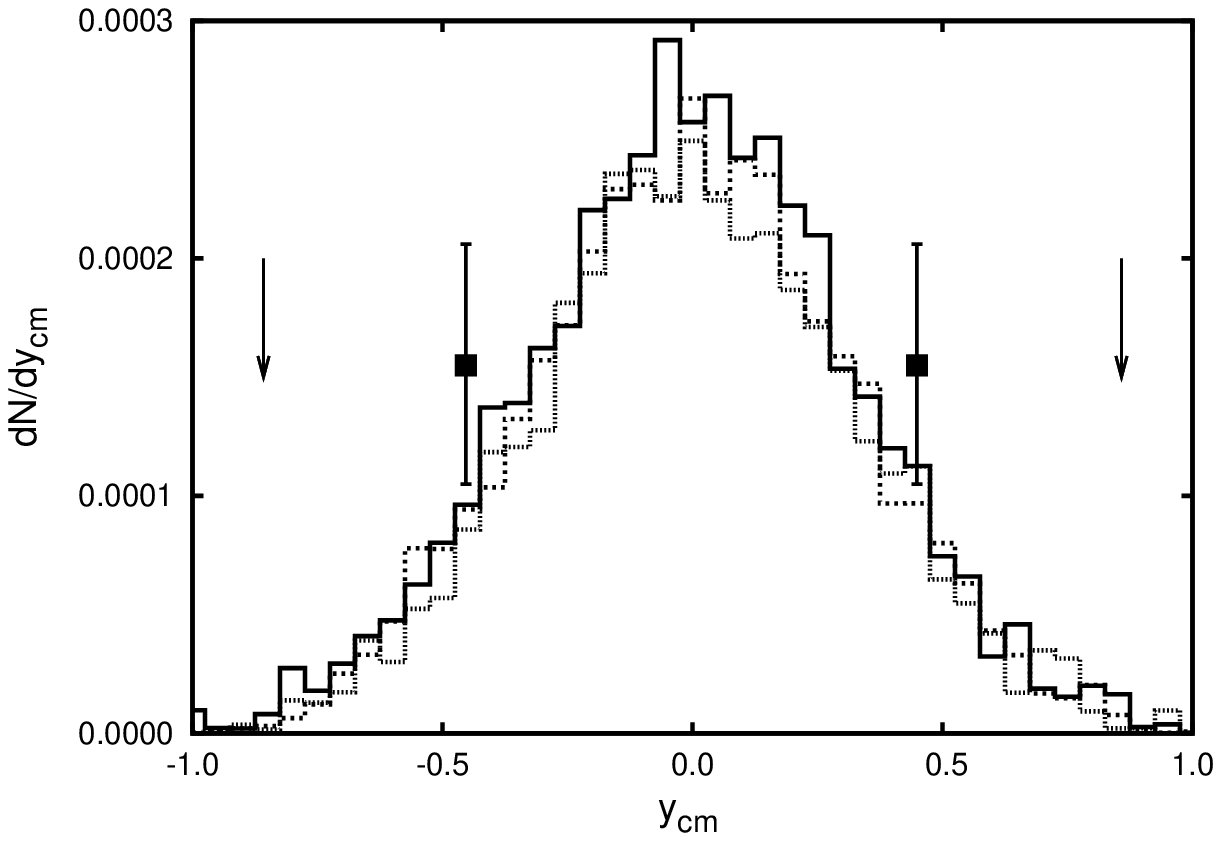}
\caption{The same as in Fig.~\ref{fig:rapkminus} but for $\phi$ mesons. The corresponding yields are $2.2 \times 10^{-4}$, $2.0 \times 10^{-4}$, and $1.9 \times 10^{-4}$ for incompressibilities $\kappa$ = 215 MeV (default), 290 MeV, and 380 MeV, respectively (solid, dashed and dotted histograms from top to bottom).}
\label{fig:rapphi}
\end{center}
\end{figure}
The transverse mass spectra and rapidity distribution for $K^-$ are exhibited in Figs.~3 and 4. Again, a reasonable agreement with data can be observed. The $m_t$ spectra and the rapidity distribution point to the tendency of slightly overestimating the data \cite{Agakishiev:2009ar}. The $K^-$ multiplicity of $7.8 \times 10^{-4}$ from our calculation is to be compared with $(7.1 \pm 1.9) \times 10^{-4}$ from data \cite{Agakishiev:2009ar}.\\
Some point of interest is the equation of state, i.e.~the incompressibility $\kappa$ of the nuclear matter, and its impact on $K^-$ yield (cf.~\cite{Aichelin:1986ss} for the impact on the $K^+$ yield). Besides the default soft equation of state ($\kappa = 215$ MeV) we display in Fig.~\ref{fig:rapkminus} the $K^-$ rapidity distributions for momentum-dependent medium ($\kappa = 290$ MeV) and hard ($\kappa = 380$ MeV) equations of state. One observes a small effect of 13 \% of the stiffness/softness of the equation of state for $K^-$ showing up as a reduction of the $K^-$ yield with increasing incompressibility. This can be explained, in line with \cite{Aichelin:1986ss}, by the fact that a stiffer mean field allows for less compression of the nuclear medium and, therefore, for fewer hard $NN$ collisions and thus not so many produced $K^-$ mesons (even in multi-step processes involving the hyperons as produced in association with $K^+$).\\
Figures 5 and 6 exhibit the $\phi$ transverse mass spectrum in the rapidity bin $0.2 < y_{lab} < 0.6$ and the $\phi$ rapidity distribution. We note the fairly good agreement with the data \cite{Agakishiev:2009ar}. Our calculated $\phi$ multiplicity is $2.2 \times 10^{-4}$, depending again on the 15 \% level on the stiffness of the equation of state, while \cite{Agakishiev:2009ar} reports a value of $(2.6 \pm 0.8) \times 10^{-4}$ from the experiment.\\
The resulting multiplicity ratio $\phi/K^-$ is 0.28, while \cite{Agakishiev:2009ar} quotes $0.37 \pm 0.13$. The slight overestimate (10 \%) of $K^-$ and the slight underestimate (15~\%) of $\phi$ in comparison with experiment is the reason for our underestimate (30 \%) of the experimental $\phi/K^-$ ratio. The $\phi/K^-$ ratio of 0.37 (0.28) translates into a contribution of 18 \% (14 \%) of the channel $\phi \rightarrow K^+ K^-$ to the $K^-$ yield. \\

\section {Detailed analysis}

After the above comparison of data and simulations for the phase-space distributions of $K^\pm$ and $\phi$, we are going to consider a few details accessible in the transport code.

\subsection{Time evolution}

\begin{figure}
\begin{center}
 \parbox{190mm}{
\includegraphics[width=8cm]{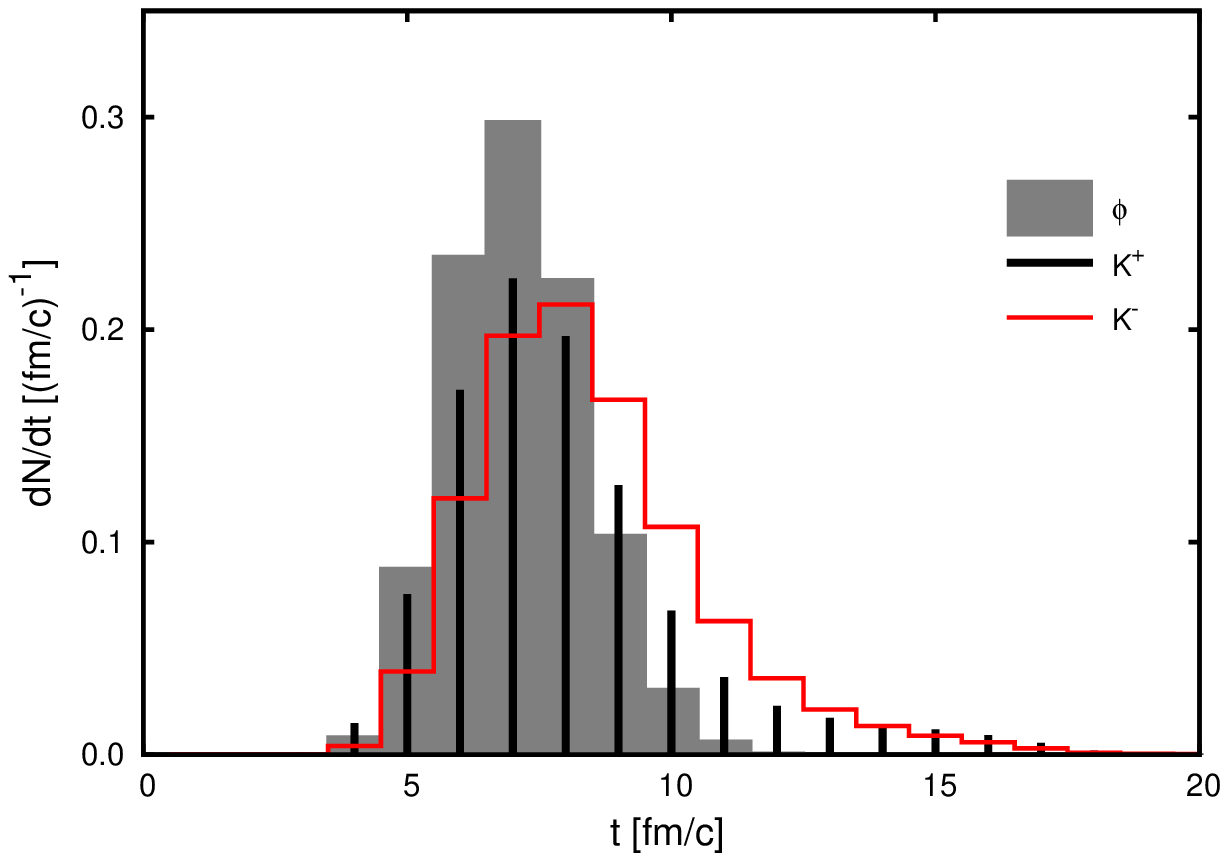}
\includegraphics[width=8cm]{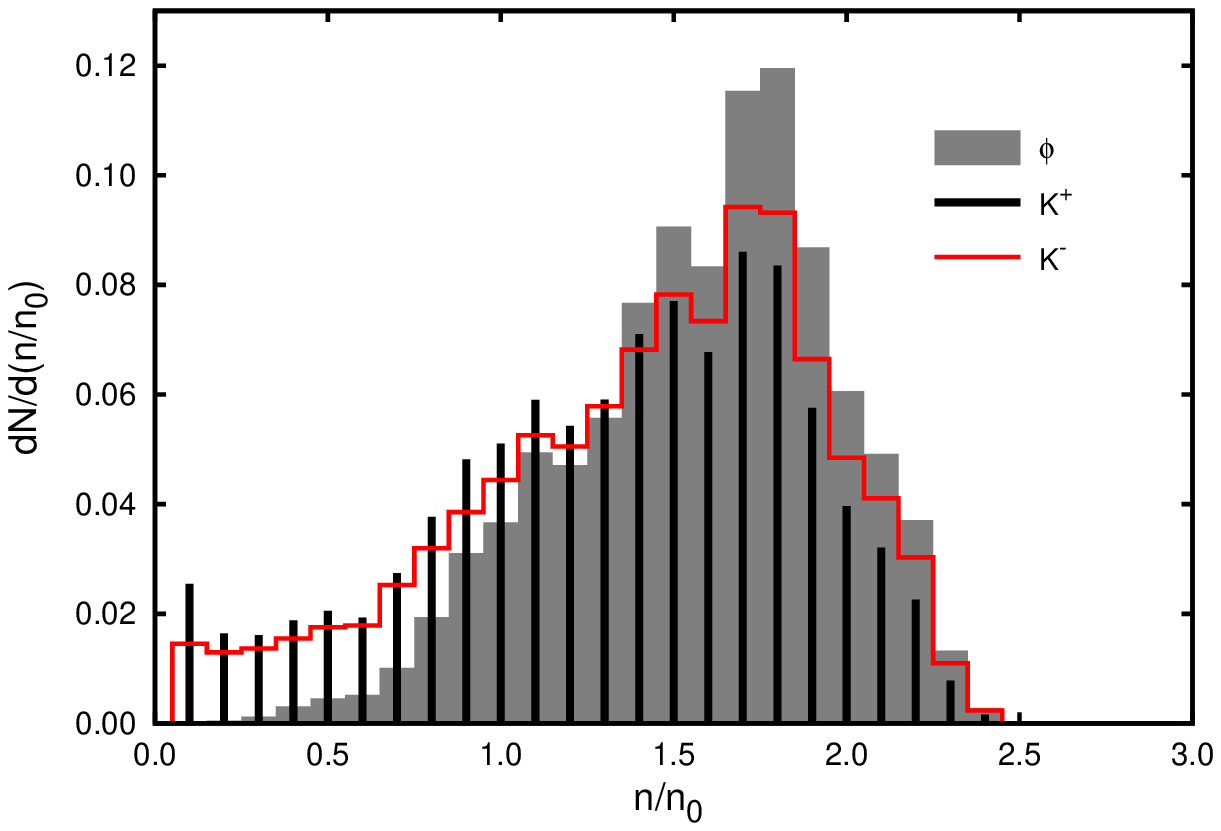}}
\caption{(color online). Creation rate of $K^+$ (solid bars), $K^-$ (histogram) and $\phi$ (grey thick bars) as a function of time (left panel) and local density (right panel).} 
\label{fig:time1density3}
\end{center}
\end{figure}

\begin{figure}
\begin{center}
 \parbox{190mm}{
\includegraphics[width=8cm]{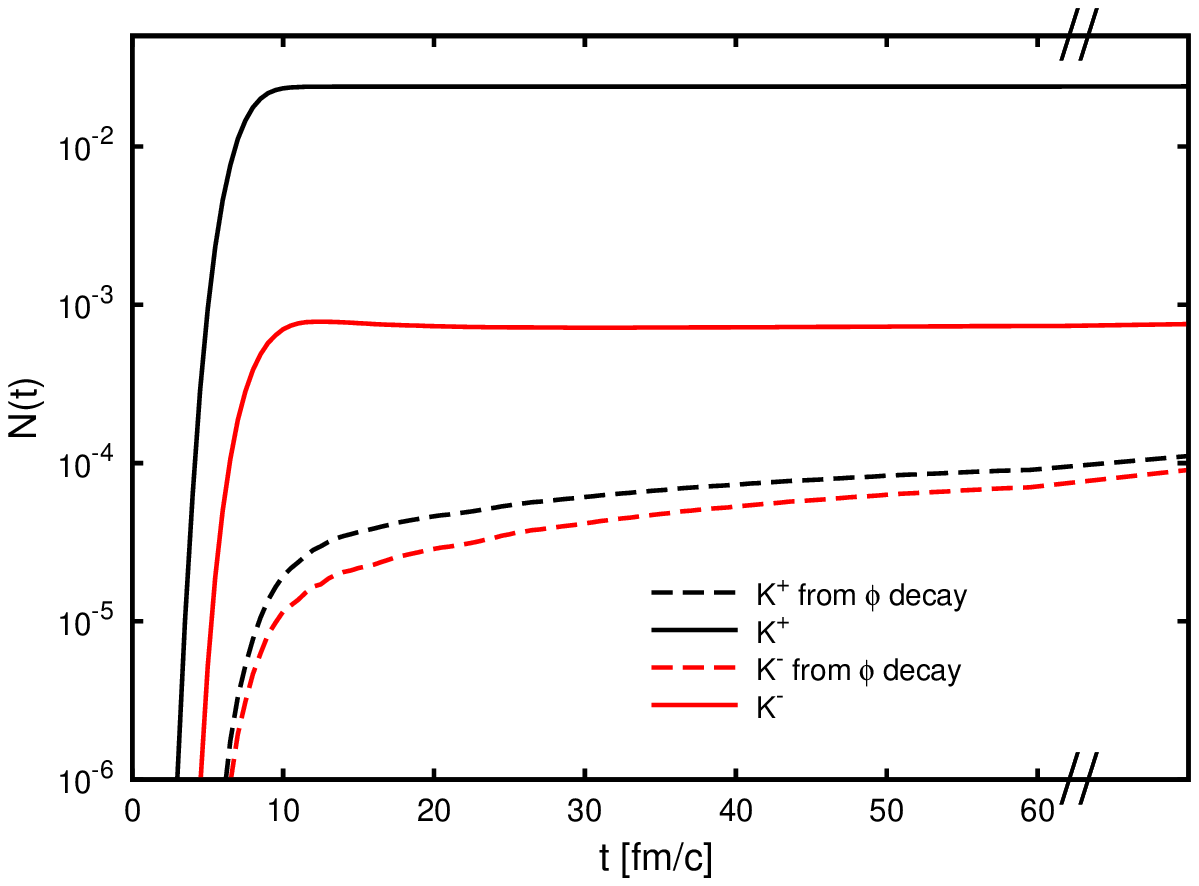}
\includegraphics[width=8cm]{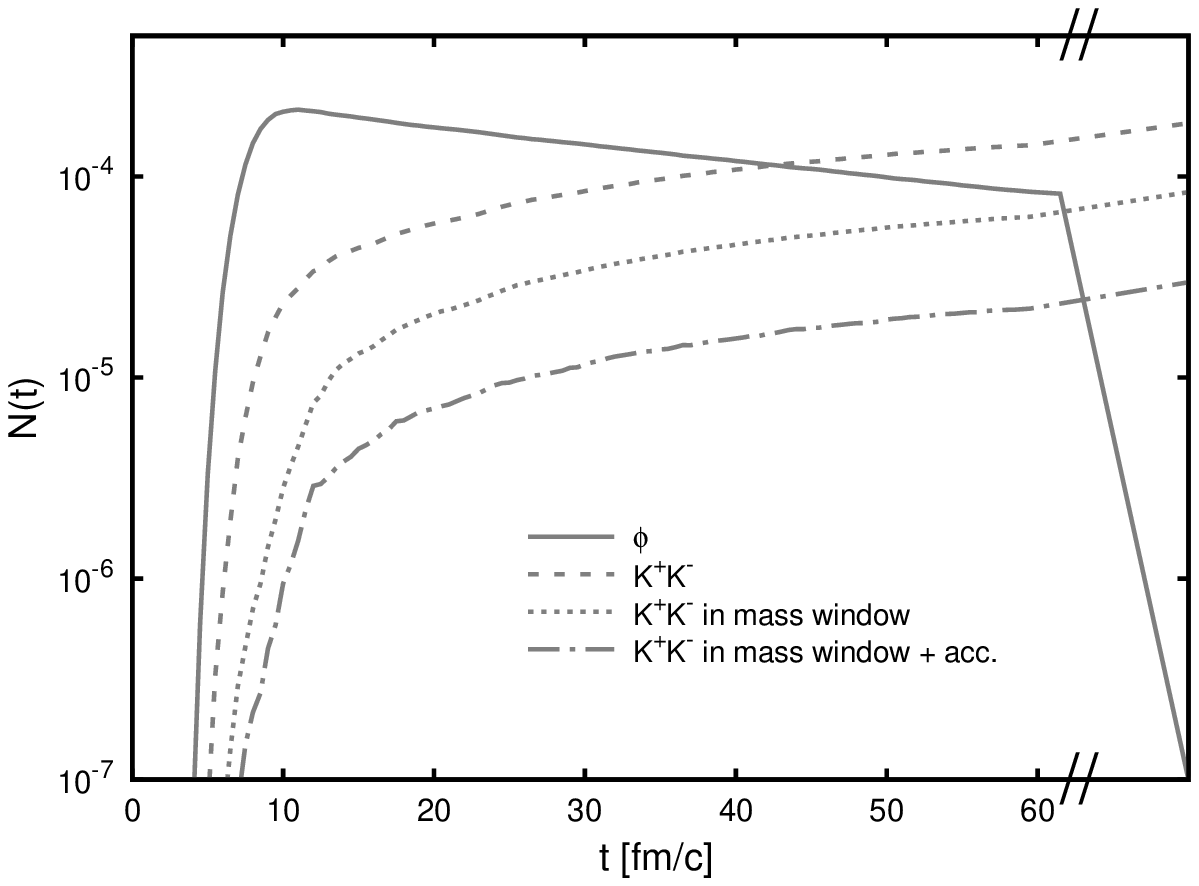}}
\caption{Left panel: (color online). Number of $K^+$ (black curves) and $K^-$ (red curves) per event as a function of time stemming from all production channels (solid curves) and from $\phi$ decays (dashed curves). The weak offset of the dashed red (black) curve at 60 fm/c shows the contribution of $\phi$ decays to the $K^-$ ($K^+$) number on a long time scale. Right panel: Number of $\phi$'s (solid curve) and resulting $K^+-K^-$ decay pairs per event as a function of time. The total number of $K^+-K^-$ pairs is depicted by the dashed curve, while the dotted curve is for the number of $K^+-K^-$ in the invariant mass window $995$ MeV $< \sqrt{(p_{K^+} + p_{K^-})^2} < 1045$ MeV. The number of $K^+-K^-$ pairs within the invariant mass window which enter the HADES acceptance region is shown by the dot-dashed curve (in-flight $K^\pm$ decays are not considered here).}
\label{fig:time54}
\end{center}
\end{figure}

\begin{figure}
\begin{center}
\centering \includegraphics[width=8cm]{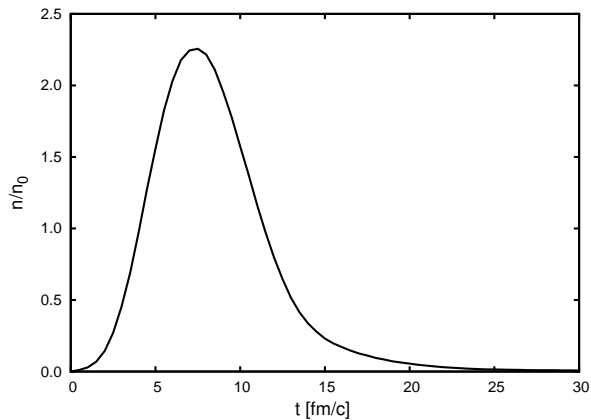}
\caption{Net baryon density in the central volume element as a function of time.} 
\label{fig:density12}
\end{center}
\end{figure}

\begin{figure}
\begin{center}
 \parbox{190mm}{
\includegraphics[width=8cm]{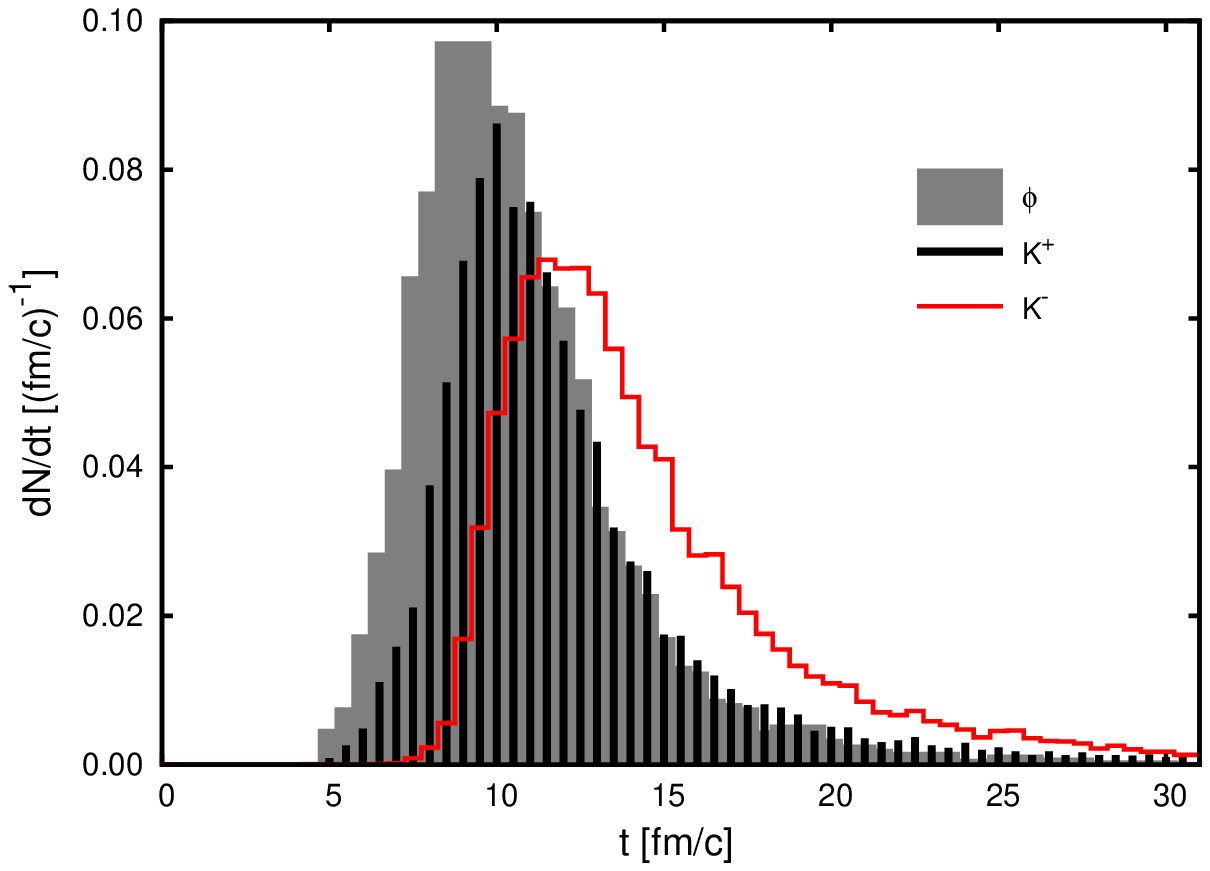}
\includegraphics[width=8cm]{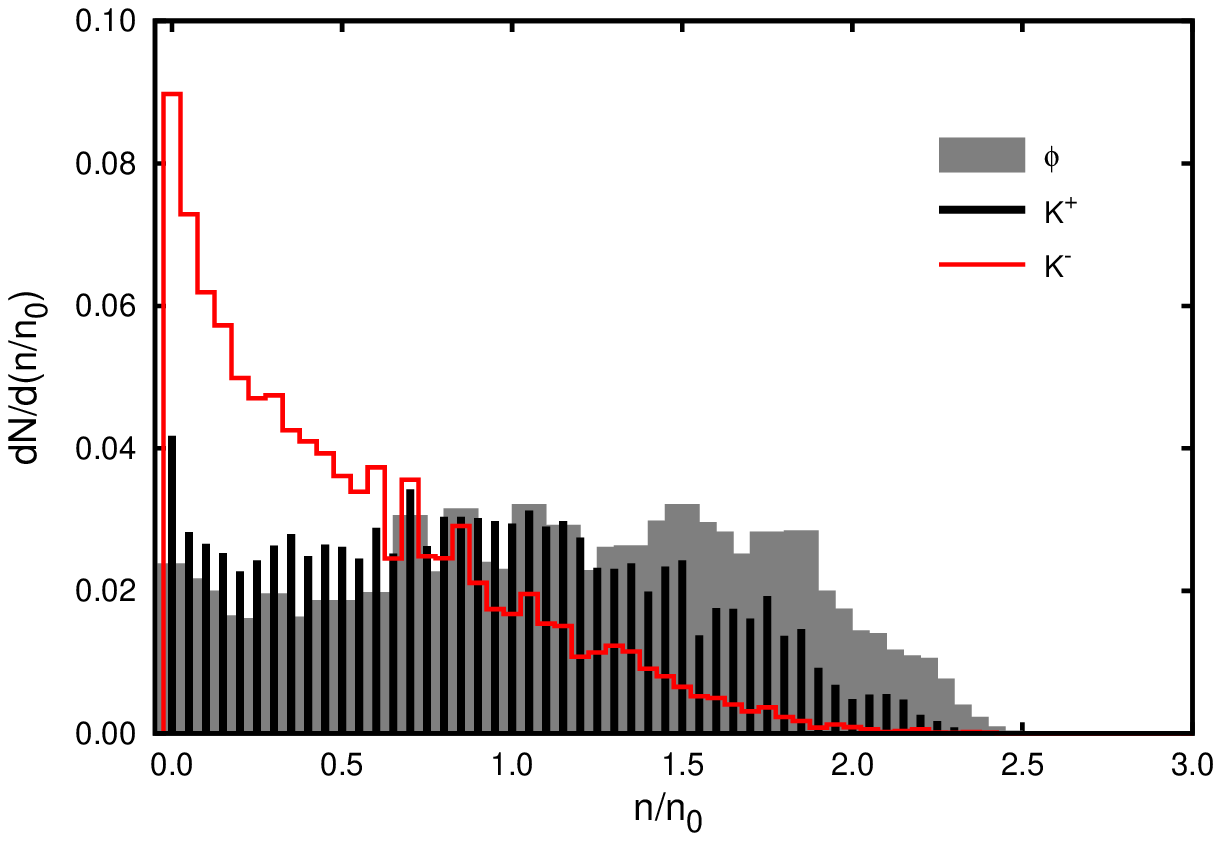}}
\caption{(color online). Distribution of last interaction as a function of time (left panel) and local density (right panel) for $K^+$ (solid bars), $K^-$ (histogram) and $\phi$ (grey thick bars).}
\label{fig:td27}
\end{center}
\end{figure}
In the left panel of Fig.~\ref{fig:time1density3} the creation rates of $K^\pm$ and $\phi$ are exhibited as a function of time. While $K^+$ and $\phi$ rates peak at the same time, the $K^-$ rate is somewhat delayed and leakes out with a longer tail. This delay is also observable in the total numbers of $K^+$ and $K^-$ as a function of time, see solid curves in left panel of Fig.~\ref{fig:time54}. (Note that, in particular for $K^-$, the absorption reactions are important for the actual total numbers.) The dotted curves in Fig.~\ref{fig:time54} (left panel) depict the $K^+$ and $K^-$ stemming from $\phi$ decays, which exhibit a stronger time lag. The $K^+$ contribution from the $\phi$ decays is negligible, of course. During the course of the collision, here followed until 60 fm/c, the contribution of $K^-$ from the $\phi$ decays  to the total $K^-$ number is below 10 \%. On a longer time scale, however, the escaping $\phi$'s decay with their vacuum branching ratio to $K^+-K^-$ thus enhancing the fraction of $K^-$ from $\phi$ decays to 14 \%. This answeres the above posed question: The $\phi$ decay channel does not dominate the $K^-$ rate, even it is not a completely insignificant channel.\\
In the experiment \cite{Agakishiev:2009ar}, the $\phi$'s are reconstructed via the requirement to have a $K^+-K^-$ pair with invariant mass in the window 995 - 1045 MeV. However, one or both of the $K^\pm$ may undergo rescattering and the final $K^+-K^-$ pair may be outside of this window. It may also happen that the $K^-$ is absorbed. Furthermore, one or both of the decay products $K^\pm$ may be outside the HADES acceptance. (The geometrical acceptance of HADES (polar angle coverage of $18^\circ$ - $85^\circ$) and hadron identification capabilities (momenta from 100 MeV/c to 1400 MeV/c) are used, cf.~\cite{Agakishiev:2009ar}.) These effects are easily accounted for in our transport code. For this purpose, the time evolution of the number of $\phi$ mesons is displayed in the right panel of Fig.~\ref{fig:time54} by the solid curve. After achieving a maximum at about 10 fm/c, the number of $\phi$'s is diminished due to decays. The dashed curve depicts these $K^+-K^-$ pairs from $\phi$ decays. Due to rescattering and/or $K^-$ absorption (which is responsible for the fact that the $K^-$ number from $\phi$ decays stay below the corresponding $K^+$ number in left panel of Fig.~\ref{fig:time54}) the number of $K^+-K^-$ pairs within the given invariant mass window is reduced by about 26 \%, both during the course of collision and the end of a long time scale. The number of $K^+-K^-$ pairs within the given invariant mass window and within the above described geometrical acceptance of HADES is again smaller by about 63 \%.\\
After the discussion of the time evolution of $\phi$ and resulting $K^+-K^-$ pairs, we briefly consider the environment of $\phi$ and $K^\pm$ production. The creation rates as a function of the local density (i.e.~in a box of 1 fm$^3$ centered at the creation point) are exhibited in the right panel of Fig.~\ref{fig:time1density3}. The $\phi$ production is clearly peaked at a density of 1.75 $n_0$, with tails from 0.3 $n_0$ to 2.3 $n_0$. Though $K^+$ and $K^-$ production happens at a noticeable extent also at lower densities, it peaks at the same local density as $\phi$ production and extending to 2.3 $n_0$. To have a guidance for relating the time evolution and the maximum achievable density, we exhibit in Fig.~\ref{fig:density12} the time evolution of the baryon density in the central volume element (1 fm$^3$). The maximum density of 2.4 $n_0$ is achieved at 7.5 fm/c.

\subsection{Freeze-out}

We turn now to the freeze-out conditions. In Fig.~\ref{fig:td27} we plot the distribution of the last contact of $K^+$, $K^-$ and $\phi$ with the ambient hadron medium as a function of time (left panel) and as a function of the local baryon density (right panel). The $\phi$'s maximum decoupling rate is somewhat earlier than the one of $K^+$. Most striking is, however, the significantly later decoupling of the $K^-$ mesons from the medium. This can be explained by the mean free paths. The $K^-$ mesons decouple later from the fireball than the $K^+$ and $\phi$ mesons as the total cross section for $K^-$ is noticeably larger causing a short mean free path. This picture fits well in the considerations in \cite{Forster:2003vc}, where the later freeze-out of $K^-$ was deduced from experimental slope parameter arguments of the transverse momentum spectra. The difference of $K^-$ and $K^+$ freeze-out is also clearly visible upon inspection of the decoupling rates as a function of local density, see right panel in Fig.~\ref{fig:td27}. $K^-$ decoupling proceeds at strikingly lower densities than $\phi$ and $K^+$ decoupling. This makes the $K^-$'s to probes for low densities below 0.5 $n_0$, in contrast to $K^+$ mesons which are sensitive to all densities, similar to the $\phi$.\\
All considerations till here are for an impact parameter of $b = 3.9$ fm adjusted to the experimental $K^+$ rapidity distribution (see Fig.~\ref{fig:rapkplus}). 

\subsection{Centrality dependence and individual channels}

\begin{figure}
\begin{center}
\centering \includegraphics[width=8cm]{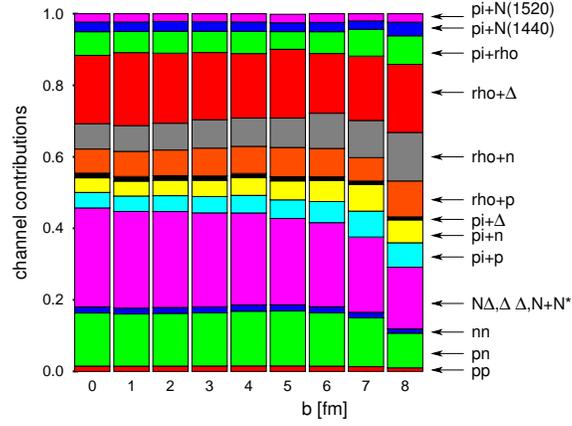}
\caption{(color online). Contribution of individual channels for $\phi$ production for various impact parameters $b$. The channels which produce a $\phi$ and two nucleons in the exit channel are $pp$, $nn$, $pn$ and $B \Delta$ (= $N\Delta$, $\Delta \Delta$ and $N$ + Resonances ($N^*$)). One nucleon and a $\phi$ are produced in $\pi +p$, $\pi +n$, $\pi +\Delta$, $\pi +N(1440)$, $\pi +N(1520)$, $\rho +p$, $\rho +n$ and $\rho +\Delta$ collisions. The channel $\pi + \rho$ produces a $\phi$.}
\label{fig:phiprod}
\end{center}
\end{figure}

\begin{figure}
\begin{center}
 \parbox{190mm}{
\includegraphics[width=8cm]{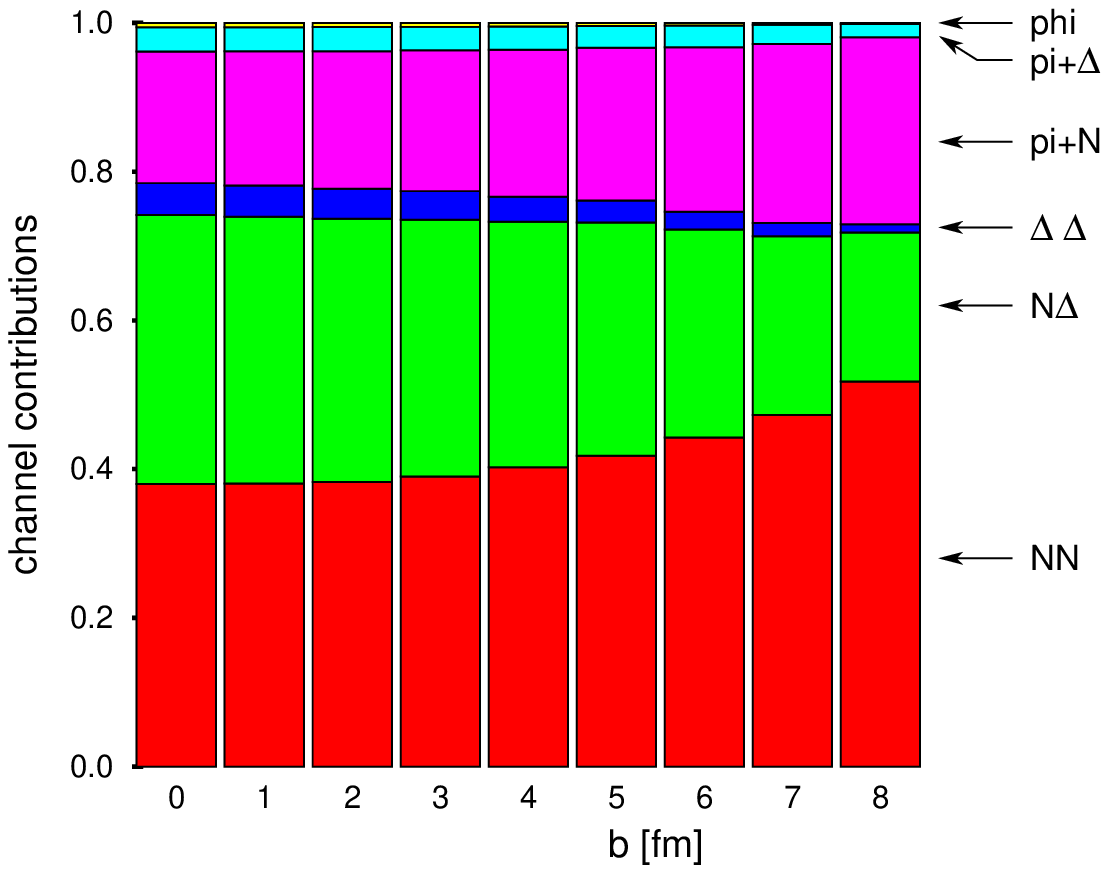}
\includegraphics[width=8cm]{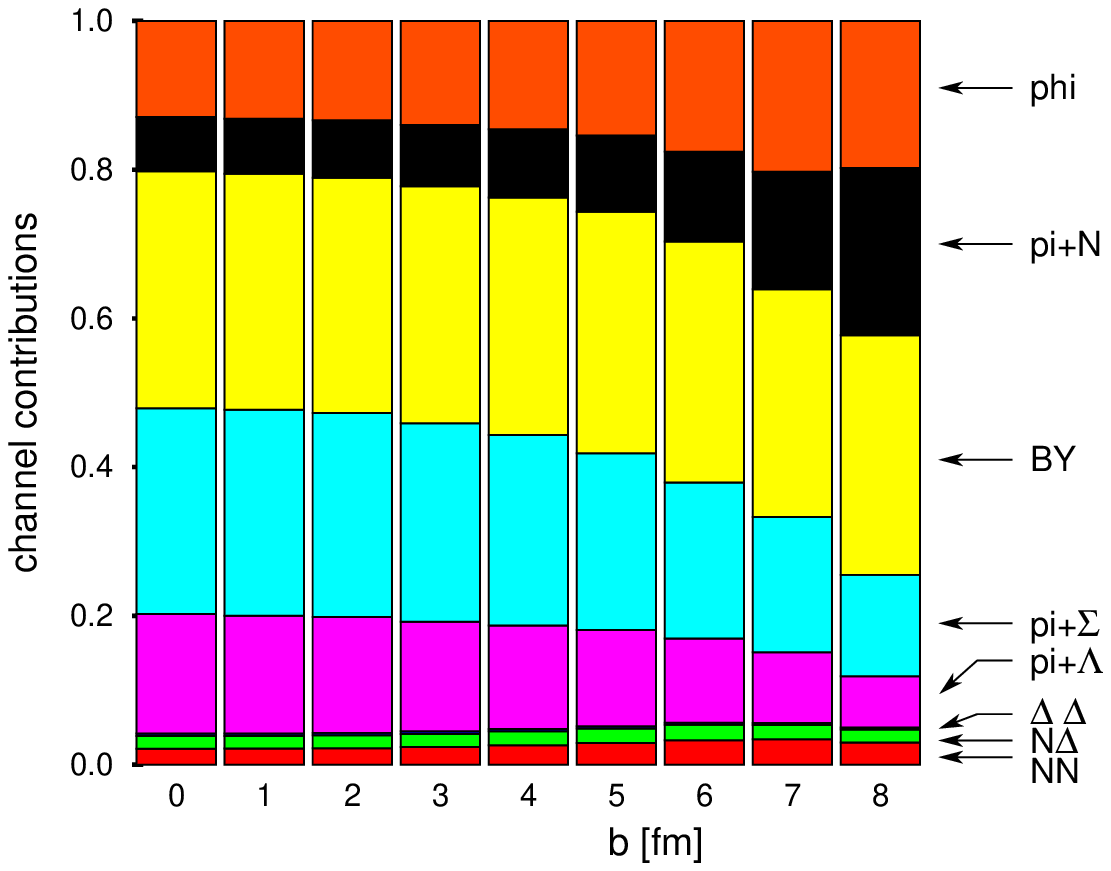}}
 \caption{(color online). The same as in Fig.~\ref{fig:phiprod} but for $K^+$ (left panel) and $K^-$ (right panel). The various channels are listed in the Appendix, cf.~Eqs.~(\ref{eq16}) - (\ref{eq23}).} 
\label{fig:kaonprod}
\end{center}
\end{figure}
Besides the beam energy dependence and the system size dependence, the centrality dependence is an important feature in heavy-ion collisions. In Fig.~\ref{fig:phiprod} the impact parameter dependence of the normalized weight of the considered channels is displayed for $\phi$ production. One observes a weak dependence on impact parameter. Most significant is, for peripheral collisions, the drop of the $NN$ and $B \Delta$ channels, whereas $\pi N$ and $\rho N$ channels expand. Clearly dominating yields are from the $NN$, $\rho N$ and $B\Delta$ channels.\\
For the $K^+$ mesons (see left panel in Fig.~\ref{fig:kaonprod}), we observe a rising (dropping) weight of the channel $NN$ ($N\Delta$) with rising impact parameter. The channels $NN$, $N\Delta$ and $\pi N$ dominate. \\
For the $K^-$ mesons (right panel of Fig.~\ref{fig:kaonprod}), most significant are the drop of the channel $\pi Y$ and a rise of the channel $\pi N$ with increasing impact parameter. Dominating channels are the strangeness transfer reactions $\pi \Lambda (\Sigma) \leftrightarrow K^- N$ in central collisions with a strong contribution of $BY \leftrightarrow NNK^-$.\\
These impact parameter dependencies can be compared with the investigation in \cite{Hartnack:2001zs}, where for the heavy collision system Au + Au at 1.5 AGeV kinetic beam energy a stronger centrality dependence of $K^\pm$ mesons has been reported. For the most central collisions the contribution of the strangeness transfer reactions $\pi \Lambda (\Sigma) \leftrightarrow K^- N$ is about 67 \% \cite{Hartnack:2001zs} (see right panel of figure 4 there) which is roughly comparable to the contribution of this channel to the total $K^-$ production yield of 42 \% which we find for Ar (1.756 AGeV) + KCl within our transport approach. Note that \cite{Hartnack:2001zs} does not include the $\phi$ channel. In a modified calculation with our code, where the $\phi$ channels are ``switched off'' we find for central collisions Au (1.5 AGeV) + Au a contribution of the strangeness-exchange channel $\pi \Lambda (\Sigma) \leftrightarrow K^- N$ to the total $K^-$ yield of 62 \%, in good agreement with \cite{Hartnack:2001zs}.

\subsection{Equation of state and effective in-medium masses}

The influence of the equation of state for $K^-$ production has been considered in Section 2, see Fig.~\ref{fig:rapkminus}. In line with the findings in \cite{Hartnack:2007wu, Hartnack:2005tr, Hartnack:2001zs, Hartnack:2003dt} we find a reduction of $K^-$ production by 11 \% when going from the soft default equation of state with incompressibility $\kappa = 215$ MeV to a hard one with $\kappa = 380$ MeV. Concerning $\phi$, a reduction of its yield by 15 \% for the hard equation of state is observable.\\
Now, we investigate the role of the hitherto assumed in-medium masses. In line with the relativistic mean field calculations in \cite{SchaffnerBielich:2000jy} we attribute schematically the interaction of $K^\pm$ and $\phi$ with ambient nucleons to an effective in-medium mass caused by a Schr\"odinger type potential. A convenient parametrization for the such modified masses is $m^*_i = m_i [1+C_i (n/n_0)]$ with $n_0$ = 0.16 fm$^{-3}$ and $C_{K^+} = +0.047$, $C_{K^-} = -0.152$ and $C_{\phi} = -0.022$ yielding mass shifts for $K^+$, $K^-$, and $\phi$ of approximately $+23$ MeV, $-75$ MeV and $-22$ MeV, respectively, at $n_0$. Putting simultaneously $C_{K^+} = C_{K^-} = C_\phi = 0$, i.e.~switching off the in-medium potentials causing effective mass shifts, we find only for $K^-$ mesons a significant effect in the rapidity distribution. The yield drops to about 60 \% without potential compared to the case with default in-medium masses. A marginal up-shifting (about 15 \%) of the yield of $K^+$ mesons is obtained, which can be explained by the smaller repulsive potential compared to the attractive $K^-N$ potential. This is an interesting result when we compare with the analysis of $K^\pm N$ potentials in \cite{Hartnack:2001zs, Hartnack:2001zs_err}. If we separately switch off the $K^-N$ and $K^+N$ potentials, a strong dependence on the $K^-N$ potential and a weak dependence on the $K^+N$ potential is observed concerning the final $K^-$ yield. This qualitatively agrees with \cite{Hartnack:2001zs_err} where, however, the reaction Au (1.48 AGeV) + Au has been analyzed. In fact, applying our code to this case we reproduce the presented peaks of $N_{K^-}(t)$ for $C_{K^+} = 0$, $C_{K^-} \neq 0$ as well as the peaks of $N_{K^-}(t)$ for the other calculations $C_{K^+} \neq 0$, $C_{K^-} \neq 0$, and $C_{K^+} = 0$, $C_{K^-} = 0$ and $C_{K^+} \neq 0$, $C_{K^-} = 0$ and their relative ordering.\\
Almost no impact (less than 10 \%) of the effective mass is observable for $\phi$ mesons.

\subsection{Comparison with FOPI data}

\begin{table}
\begin{center}
\begin{tabular}{l|c}
  yields from & Ni (1.93 AGeV) + Ni\\
  \hline
  $B$ + $B$  		& $11.2 \times 10^{-4}$\\
  $\pi$ + $B$  		& $2.4 \times 10^{-4}$\\
  $\rho$ + $B$ 		& $8.6 \times 10^{-4}$\\
  $\pi$ + $\rho$ 	& $1.5 \times 10^{-4}$\\
  $\pi$ + $N(1440)$ 	& $0.6 \times 10^{-4}$\\
  $\pi$ + $N(1520)$ 	& $0.5 \times 10^{-4}$\\
  \hline
  total yield 		& $2.5 \times 10^{-3}$ \\
  \hline	
  data from experiment \cite{Mangiarotti:2003es}\\
  for $T$ = 130 MeV 		& $(1.2\pm 0.4\pm 0.6) \times 10^{-3}$\\
  for $T$ = 70 MeV 		& $(4.5\pm 1.4\pm 2.2) \times 10^{-3}$\\
\end{tabular}
\caption{Comparison of the yields from various channels (see text for explanations) and comparison with FOPI data \cite{Mangiarotti:2003es} for two different assumptions for the slope parameter $T$ of $\phi$ phase-space distribution.} 
\label{tab:FOPI}
\end{center}
\end{table}

\begin{figure}
\begin{center}
\centering \includegraphics[width=8cm]{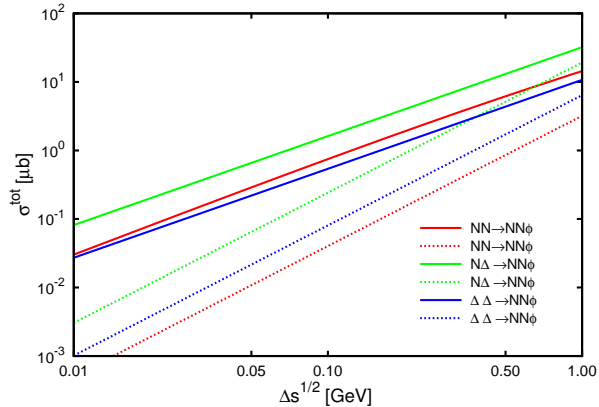}
\caption{(color online). Compilation of isospin-averaged cross sections for the production of $\phi$ mesons in baryon-baryon collisions as a function of the excess energy. The dotted curves represent the parametrization of \cite{Chung:1997mp}, while the solid curves are for the parametrization of \cite{Kaptari:2004sd, Kaptari:2008nb} for various reactions indicated in the legend.} 
\label{fig:sigma1}
\end{center}
\end{figure}

\begin{figure}
\begin{center}
 \parbox{190mm}{
\includegraphics[width=8cm]{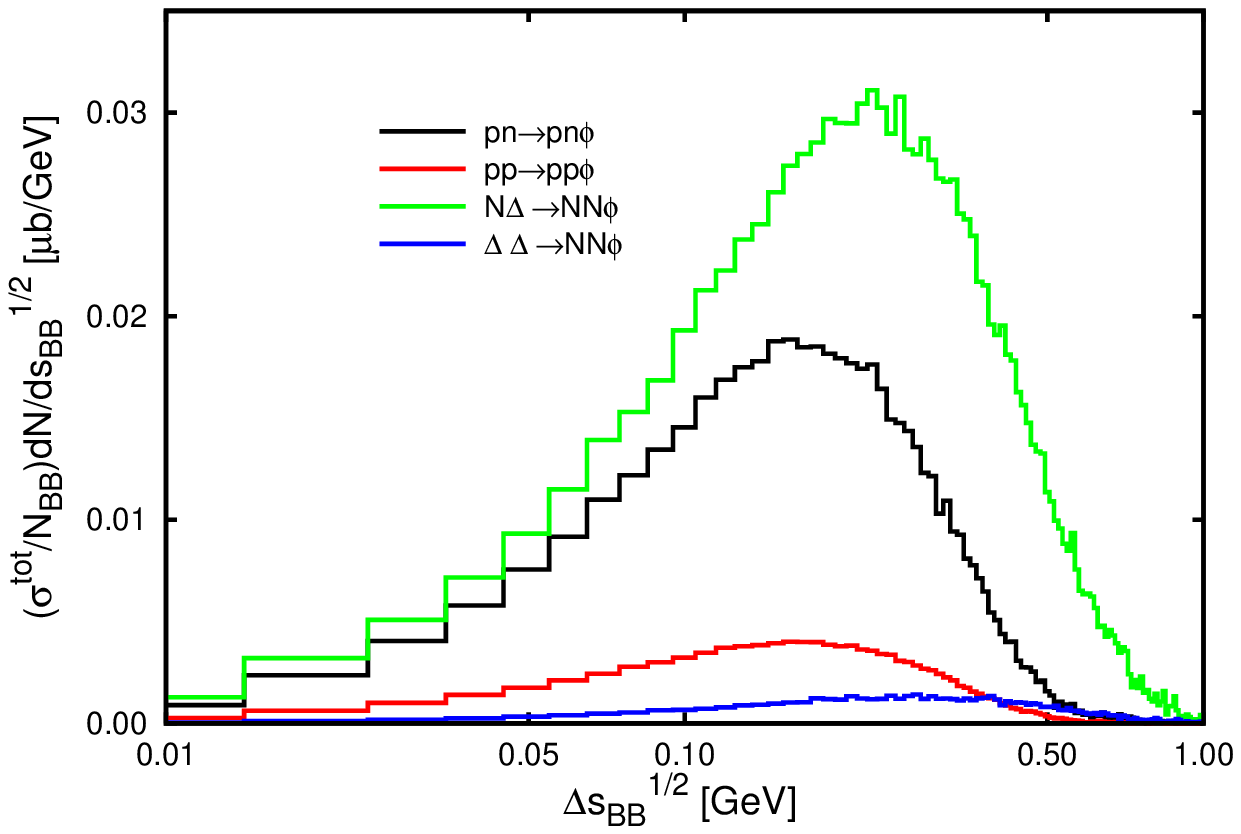}
\includegraphics[width=8cm]{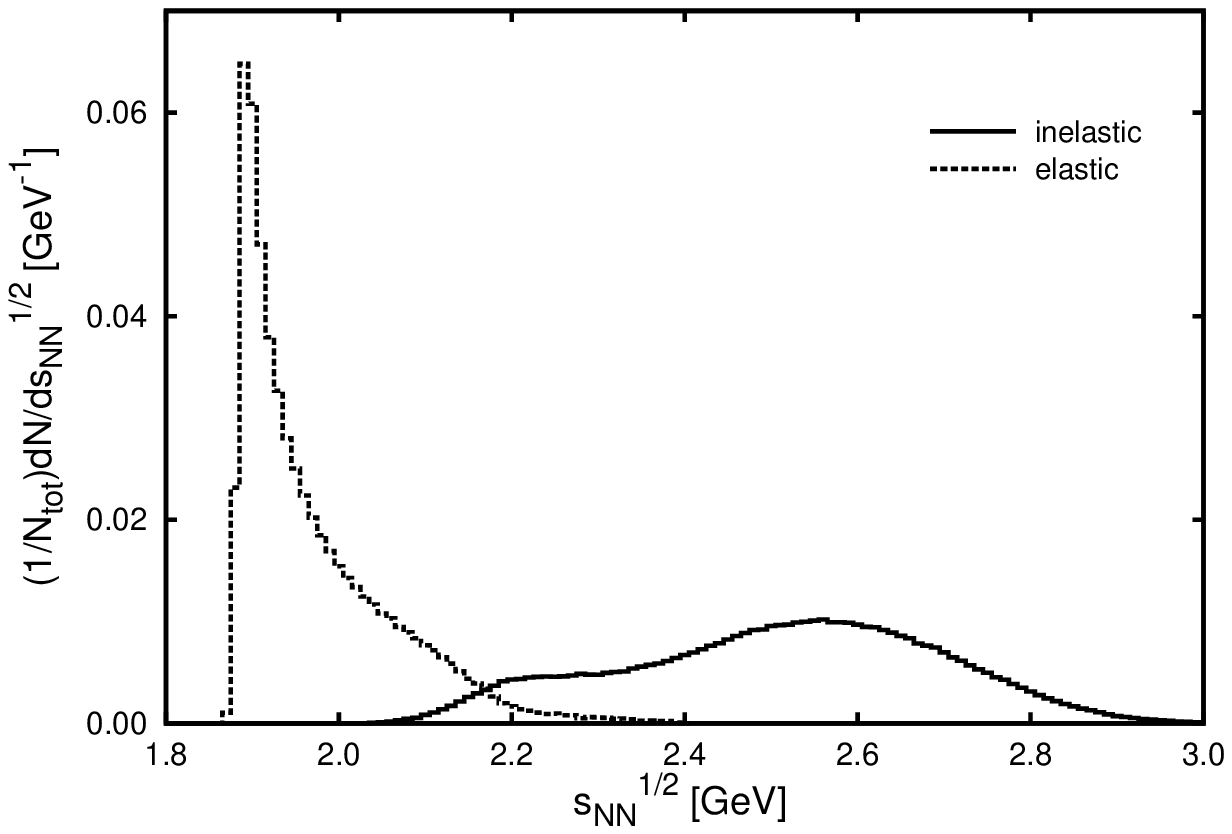}}
\caption{Left panel: (color online). Normalized $\phi$ production distribution of various isospin-dependent baryon-baryon reaction channels as a function of the excess energy $\Delta s^{1/2}_{BB}$ with cross sections from \cite{Kaptari:2004sd, Kaptari:2008nb} whereas $\sigma_{pp\rightarrow pp\phi} \equiv \sigma_{nn\rightarrow nn\phi}$. Right panel: Normalized distribution of elastic (dashed curve) and inelastic (solid curve) $NN$ collisions as a function of $\sqrt{s_{NN}}$. For the reaction Ar (1.756 AGeV) + KCl.}
\label{fig:sigma23}
\end{center}
\end{figure}
The above analysis focuses entirely on the new HADES data \cite{Agakishiev:2009ar} for Ar (1.756 AGeV) + KCl collisions. To check the consistency with the previous FOPI data \cite{Mangiarotti:2003es} for the reaction Ni (1.93 AGeV) + Ni and the first attempt \cite{Barz:2001am} for their interpretation we calculate the $\phi$ yields. The results are listed in Tab.~\ref{tab:FOPI} with $B = N$, $\Delta$. The individually specified channels producing $\phi$'s are baryon-baryon ($B$ + $B$), pion-baryon ($\pi$ + $B$), rho-baryon ($\rho$ + $B$), pion-rho ($\pi + \rho$) and pion-baryon resonances ($\pi$ + $N(1440)$ and $\pi$ + $N(1520)$) collisions. This sorting enables an easy comparison with the corresponding table~2 in \cite{Barz:2001am}. We emphasize that we employ throughout the paper the new parametrization \cite{Kaptari:2004sd} for the nucleon-nucleon channel $NN \rightarrow NN\phi$ which agrees well with available data in the $pp \rightarrow pp \phi$ channel (see \cite{Kaptari:2008nb, Maeda:2007cy}). A comparison of this parametrization according to \cite{Kaptari:2004sd, Kaptari:2008nb} with the previously used \cite{Barz:2001am} from \cite{Chung:1997mp} is exhibited in Fig.~\ref{fig:sigma1}. One recognizes large differences near the threshold. The relative strengths of $NN$, $N\Delta$ and $\Delta\Delta$ channels is in both parametrizations the same. The number of baryon-baryon collisions which may create a $\phi$ at given excess energy is in fact much larger for small excess energies. However, due to the weighting with corresponding cross sections, the main number of $\phi$'s created in baryon-baryon collisions is at excess energies of 200 - 300 MeV (see left panel of Fig.~\ref{fig:sigma23}). The $N\Delta$ channel followed by the $pn$ channel clearly dominate.\\ 
These new $pp\rightarrow pp\phi$ and $pn\rightarrow pn\phi$ cross sections enhance significantly the yields in comparison with the previous analysis \cite{Barz:2001am}. One observes that the FOPI data \cite{Mangiarotti:2003es}, which are extrapolated from small phase-space pieces to $4\pi$, are also nicely reproduced within the present approach. Direct baryon-baryon collisions represent the dominating channel followed by rho-baryon collisions; pion-induced $\phi$ production in individual channels resolved in Tab.~\ref{tab:FOPI} is subleading.\\
For the sake of completeness we depict in the right panel of Fig.~\ref{fig:sigma23} the distribution of elastic $NN$ collisions and a selection of inelastic $NN$ collisions as a function of $\sqrt{s_{NN}}$. This analysis is based on the channels $NN\rightarrow NN$, $NN\rightarrow N$ + Resonances, $NN\rightarrow N\Delta$, $NN\rightarrow \Delta\Delta$ and the direct one-pion production channel $NN\rightarrow NN\pi$.

\section{Summary}

In summary we apply a transport code of BUU type to analyze the new experimental data of inclusive $K^\pm$ and $\phi$ production in the reaction Ar (1.756 AGeV) + KCl. With a given parameter set (in-medium masses parametrizing the influence of the ambient baryon medium on spectral properties of hadrons, mean field, equation of state, and individual cross sections) the available data \cite{Agakishiev:2009ar} are well described and the good agreement with previous data \cite{Mangiarotti:2003es} for the reaction Ni (1.93 AGeV) + Ni is ensured. We note the tendency to reproduce the yields reported in \cite{Mangiarotti:2003es, Agakishiev:2009ar} in the lower parts of the error intervals. The improved $NN \rightarrow NN \phi$ cross section from \cite{Kaptari:2004sd, Kaptari:2008nb} is essential to achieve such an agreement (cf.~\cite{Barz:2001am}). With respect to the situation reported in \cite{Barz:2001am} and the question posed in \cite{Kampfer:2001mc} (i.e.~how important is the feeding from $\phi$ decays for the $K^-$ yield) the role of the channel $\phi \rightarrow K^+ K^-$ has been clarified: While the astonishingly large ratio of $\phi/K^-$ is reproduced within error bars, the decay channel $\phi \rightarrow K^+K^-$ is nevertheless subleading for $K^-$ production. Most of $\phi$'s decay, of course, outside of the compressed nuclear medium so that the anticipated understanding \cite{Hartnack:2007wu, Hartnack:2005tr, Hartnack:2001zs, Hartnack:2001zs_err, Hartnack:2003dt} of $K^-$ dynamics in relativistic heavy-ion collisions gets only minor modifications by the $\phi$ channel.\\
We find that the strangeness transfer reactions $\pi Y \leftrightarrow NK^-$ and $BY \leftrightarrow NNK^-$ yields about 42 \% and 32 \% of the finally observed $K^-$ in the considered collisions. Nevertheless, for a profound understanding of the strangeness dynamics the contribution of $\phi$ dynamics is not negligible. Obviously, there is still space for improvements; may be the catalytic reactions of \cite{Kolomeitsev:2009yn} serve to this. Further forthcoming data from heavier collision systems envisaged after the HADES upgrade will shed light on such details.\\
Freezing in all other parameters we find a decrease of $\phi$ yield by 15 \% when we use a hard equation of state instead of a soft one; the assumed effective in-medium $\phi$ mass increases the $\phi$ yield by less than 10 \% in comparison with simulations employing the vacuum mass. We emphasize that in our present treatment the open and hidden strangeness mesons are considered as good (narrow) quasi-particles. With respect to hints of a considerable broadening of the $\phi$ in the nuclear medium \cite{Djalali:2008zza, Vujanovic:2009wr}, in future investigations the propagation of spectral functions should be implemented. (The experience in \cite{Barz:2006sh}, however, let us argue that the gradient expansion of Kadanoff-Baym's quantum kinetic equations results in a rapid approach towards vacuum spectral functions with small observable effects of the in-medium broadening, at least in small collision systems.) \\
Many of the elementary cross sections for $\phi$ production and $\phi$-nucleon interactions, in particular the important channels of nucleon-$\Delta$ and nucleon-$N^*$ as well as $\rho$-$\Delta$, are still purely known. Nevertheless, the satisfactory description of the data \cite{Mangiarotti:2003es, Agakishiev:2009ar} let us believe to have in total a reasonable control of the $\phi=s\bar{s}$ dynamics. This may be compared in future investigations with the dynamics of real multistrange hadrons such as the $\Xi^-$. \\
We note finally that a monitoring of the $\phi$ dynamics in the $e^+e^-$ channel is an interesting feature for further experimental studies with HADES and accompanying theoretical interpretations for which the present investigation can serve.\\
\textit{Acknowledgements:} Useful discussions with Drs.~R.~Kotte and H.~Oeschler are gratefully acknowledged. The work is supported by BMBF 06DR136 and 06DR9059D and OTKA T48833 and T71989.

\section*{Appendix}

The evolution of single-particle phase-space distribution functions $f_i(\vec r,\vec p,t)$ can be described via a set of coupled Boltzmann-\"Uhling-Uhlenbeck equations in the quasi-particle limit
\begin{equation}
\frac{\partial f_i}{\partial t}+\frac{\partial H_i}{\partial \vec p}\frac{\partial f_i}{\partial \vec r}
- \frac{\partial H_i}{\partial \vec r}\frac{\partial f_i}{\partial \vec p} = \sum_j \cal{C}_{\textit{ij}}
\label{eq1}
\end{equation}
with an approximated Hamiltonian in the local rest frame 
\begin{equation}
H_i \approx \sqrt{m_{i}^2+\vec p\;^2}+U_i.
\label{eq2}
\end{equation}
Here, $m_{i}$ defines the rest mass in a scalar momentum and density dependent mean field $U_i$ 
\begin{equation}
U_i = A \frac{n}{n_0} + B \left( \frac{n}{n_0}\right) ^\tau + C \frac{2}{n_0} \int \frac{d^3p'}{(2\pi)^3} \frac{f_i(\vec r,\vec p\;')}
{1+(\frac{\vec p - \vec p\;'}{\Lambda})^2}
\label{eq3}
\end{equation}
for hadron specie $i$, while $n$ and $n_0$ stands for the baryon number density and the normal nuclear matter density, respectively. A special type of the nucleon mean field is specified within a set of the following parameters ($A$, $B$, $C$, $\Lambda$, $\tau$) = (-0.120 GeV, 0.151 GeV, -0.065 GeV, 2.167 GeV, 1.230) representing a soft momentum-dependent equation of state (EoS) with incompressibility modulus $\kappa = 215$ MeV as default. Further parameter sets yielding an hard EoS ($\kappa = 380$ MeV) and a medium EoS ($\kappa = 290$ MeV) are also investigated.\\
The set of evolution equations (\ref{eq1}) is solved by the test-particle parallel-ensemble method (cf.~\cite{Kolomeitsev:2004np, Barz:2001am, Barz:2003wz, Barz:2006sh} for details and \cite{Bertsch:1988ik, Wong:1982zzb} for the general technique). The code handles, besides the propagation of nucleons, 24 $\Delta$ and nucleon resonances and $\pi, \eta, \omega, \rho$ and $\sigma$, the strange baryons $\Lambda$ and $\Sigma$ as well as mesons with strangeness content ($\phi$ and $K^\pm$). Hyperons experience 2/3 of the nucleon mean field.\\
Elastic and inelastic interaction between two hadrons are represented by the r.h.s.~of Eq.~(\ref{eq1}), the collisional integral $\cal{C}_{\textit{ij}}$, which embodies the \"Uhling-Uhlenbeck terms responsible for the Pauli-blocking in the collision of spin-1/2 hadrons. Particle annihilation and creation processes are realized via various $\sqrt s$-dependent cross sections.\\ 
The considered strange particles are surrounded by the nuclear medium when they are created and propagate. This leads to various meson-nucleon potentials translating into different effective $K^\pm$ and $\phi$ masses described in subsection 3.4. The nuclei are initially displaced so that the touching time for zero impact parameter is about 4 fm/c.\\
The following reactions for $K^\pm$ production are included:\\

1) baryon - baryon
\begin{equation}
NN \rightarrow \left\{ 
\begin{aligned}
 & NNK^+K^-\\
 & N Y K^+\\ 
 & \Delta Y K^+
\end{aligned} \right.
, 
N\Delta \rightarrow \left\{ 
\begin{aligned}
 & NNK^+K^-\\
 & N\Delta K^+K^-\\ 
 & N Y K^+\\
 & \Delta Y K^+
\end{aligned} \right.
, 
\Delta\Delta \rightarrow \left\{ 
\begin{aligned}
 & NNK^+K^-\\
 & \Delta \Delta K^+K^-\\
 & N Y K^+\\
 & \Delta Y K^+ ,
\end{aligned} \right.
\label{eq16}
\end{equation}

2) pion - baryon 
\begin{equation}
\pi N \rightarrow \left\{ 
\begin{aligned}
 & NK^+K^-\\
 & Y K^+ 
\end{aligned} \right.
, 
\pi \Delta \rightarrow \left\{ 
\begin{aligned}
 & NK^+K^-\\
 & Y K^+ ,
\end{aligned} \right.
\label{eq17}
\end{equation}

3) phi decay channel
\begin{equation}
\begin{aligned}
& \phi \rightarrow K^+K^- .
\end{aligned}
\label{eq21}
\end{equation}

Besides the associated production of $K^+-K^-$ and final states with hyperons (with exception of $\pi \Delta$ channel) in Eqs.~(\ref{eq16}) - (\ref{eq21}) we additionally use the following strangeness-exchange reactions for $K^-$ production:\\

4) baryon - hyperon 
\begin{equation}
\left. 
\begin{aligned}
& NY \\
& \Delta Y 
\end{aligned} \right\rbrace 
\leftrightarrow NNK^- ,
\label{eq22}
\end{equation}

5) pion - hyperon
\begin{equation}
\begin{aligned}
& \pi Y \leftrightarrow K^-N .
\end{aligned}
\label{eq23}
\end{equation}
These channels are described in \cite{Barz:2003wz} with the exception of the channel $NN \rightarrow NN\phi$ for which a new parametrization, described in subsection 3.5, is employed.

\end{document}